\documentclass[11pt]{article}
\usepackage{graphicx,moriond}

\def\be{\begin{equation}}
\def\ee{\end{equation}}
\def\bea{\begin{eqnarray}}
\def\eea{\end{eqnarray}}

\begin{document}
\title{ANISOTROPIC FLOW AT THE RELATIVISTIC HEAVY ION COLLIDER}
\author{RAIMOND SNELLINGS}
\address{
NIKHEF, Kruislaan 409, 1098 SJ Amsterdam, The Netherlands \\
E-mail: Raimond.Snellings@nikhef.nl}
\maketitle

\abstracts{
Anisotropic flow is recognized as one of the main observables providing
information on the early stage of a heavy-ion
collision~\cite{Voloshin:2002wa}. At RHIC the observed strong
collective flow of the bulk matter is considered evidence for an early
onset of thermalization, and an ideal hydrodynamical expansion with an
equation of state consistent with that obtained from lattice QCD
calculations. 
The large collective flow of the bulk and the
inferred large energy loss of the produced jets propagating through the
created matter are key signatures for the formation of a Quark Gluon
Plasma~\cite{Gyulassy:2004zy}.
}

\section{Introduction}


QCD calculations predict that a sufficient large system heated to
a temperature of approximately 170~MeV will undergo a phase transition
from normal nuclear matter to a Quark Gluon Plasma (QGP).
Heavy-ion collisions are expected to provide the best controlled
environment to create and study such a large high temperature system. 
Experiments at the Relativistic Heavy Ion Collider (RHIC), the first
heavy-ion collider, have provided a new era in the study of QCD
matter~\cite{Jacobs:2004qv}. 

\section{Azimuthal correlations versus the reaction plane}\label{sec:corr}
\subsection{Particles with intermediate and high transverse momenta}
\label{subsec:inter}

One of the most promising observables, discovered at RHIC, 
is the suppression~\cite{Vitev:2004bh} of particles with large
transverse momenta.
The predicted mechanism for this suppression, the so called
jet-quenching, is parton energy loss by induced gluon radiation. 
The magnitude of the energy loss depends on the parton density (mostly
gluons) of the created system and its size.

\begin{figure}[t]
  \begin{center}
    \includegraphics[width=.8\textwidth]{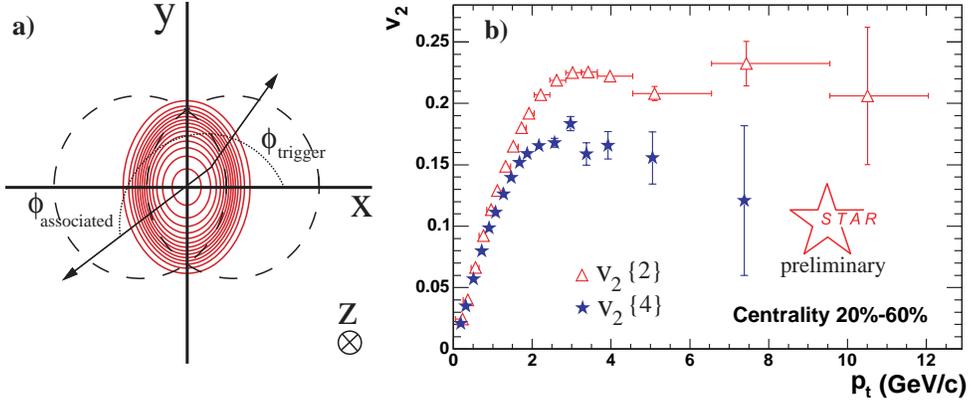}
  \end{center}
  \caption{
    a) Illustration of the nuclear overlap region in non-central heavy-ion
    collisions.
    b) $v_2$ obtained from two, $v_2\{2\}$, and four particle,
    $v_2\{4\}$, cumulant methods versus transverse
    momentum~\protect\cite{Tang:2004je}.   
  }
  \label{fig:geometry}
\end{figure}

In non-central heavy-ion collisions the nuclear overlap region in the
transverse plane has an almond like shape, see Fig.~\ref{fig:geometry}a.
In the case of parton energy loss, the particle yield
at large transverse momenta due to the spatial anisotropy of the
created system will have an azimuthal correlation with respect to the 
reaction plane (the plane spanned by the beam axis {\bf z}, and the
impact parameter, which is along the {\bf x}-axis). 
As the initial spatial geometry of the collisions is known 
the azimuthal dependence of the high-$p_t$ particle yield is
a sensitive probe to the details of jet quenching.  
The particle yield as a function of azimuthal angle can be described
by:
\[
E \frac{{\rm d^3}N}{{\rm d^3}p} = \frac{1}{2\pi} \frac{{\rm d^2}N}
{p_t {\rm d}p_t {\rm d}y}[1 + \sum_{n=1}^{\infty} 2 v_n {\rm cos}(n\phi)]
\]
The coefficient of the second harmonic of this Fourier decomposition,
$v_2$, is called elliptic flow.

\begin{figure}
  \begin{center}
    \includegraphics[width=.85\textwidth]{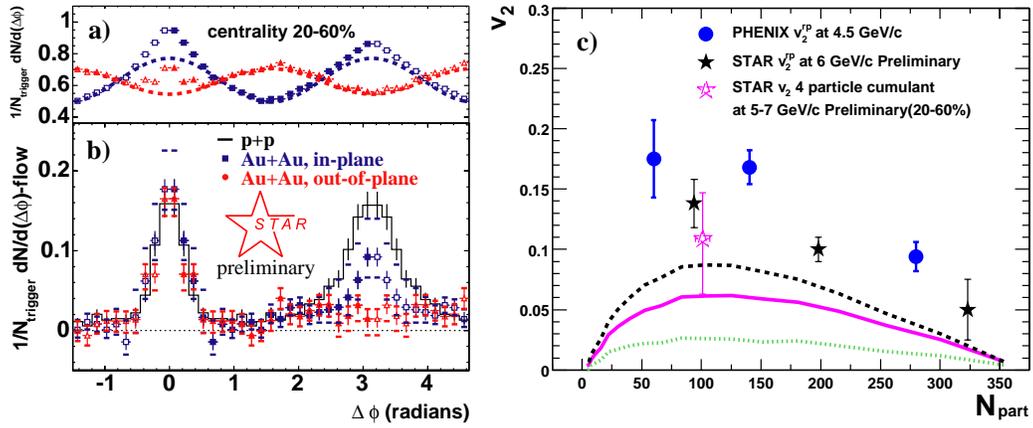}
  \end{center}
  \caption{
    a) The di-hadron azimuthal correlation in and out of the reaction
    plane 
    b) The di-hadron azimuthal correlation after subtracting the
    elliptic flow contributions
    c) $v_2$ at large transverse momenta compared to different energy
    loss scenarios~\protect\cite{Drees:2003zh}.
    \label{fig:chargedflow}}
\end{figure}

Figure~\ref{fig:geometry}b shows the elliptic flow as a function of
transverse momenta.
The elliptic flow is calculated using the two, $v_2\{2\}$, and four
particle, $v_2\{4\}$, cumulant methods~\cite{Borghini:2001vi} 
which give, as shown in
Fig.~\ref{fig:geometry}b, about 20\% difference in magnitude of
$v_2$. 
This is understood because 
the azimuthal correlation between the particles, used to calculate
$v_2$, is not completely caused by their correlation with the reaction
plane. 
Azimuthal correlations from e.g. jets and resonances, would affect the
two particle cumulant more than the higher order ones. Fluctuations
also affect $v_2\{2\}$ and $v_2\{4\}$ but when small in the opposite
direction. In our current understanding, the best estimate of the true
elliptic flow is between $v_2\{4\}$ and ($v_2\{2\} +
v_2\{4\}$)/2~\cite{Miller:2003kd}. 
The elliptic flow shown has its maximum around 3~GeV/$c$, and even at 8
GeV/$c$ is still quite substantial. In case of jet quenching the
elliptic flow is expected to be non-zero. 

The di-hadron azimuthal correlation is expected, due to the spatial
anisotropy of the system, to be sensitive to the jet quenching.
Figure~\ref{fig:chargedflow}a shows the azimuthal correlations of two
charged particles, were the trigger particle is defined as 4 $< p_t^{trig}
< 6$ GeV/$c$ and the {\it associated} particle satisfies 2 GeV/c $<
p_t < p_t^{trig}$. 
The squares show the di-hadron correlation when the
trigger particle is within $\pi/4$ aligned with the reaction plane,
the {\bf x}-axis in Fig.~\ref{fig:geometry}a. 
The stars show the di-hadron correlation when the
trigger particle is within $\pi/4$ perpendicular to the reaction
plane, the {\bf y}-axis. The dominant correlation is the opposite
oscillation due to the elliptic flow of the trigger and associated
particle, which is shown as the dashed lines in
Fig.~\ref{fig:chargedflow}a. 
In Fig.~\ref{fig:chargedflow}b the
di-hadron correlation, in and out of the reaction plane, is shown after
the elliptic flow contribution is
subtracted~\cite{Tang:2004vc,Filimonov:2004qz,Adams:2004wz,Bielcikova:2003ku}.
The near-side jet-like 
correlation in and out of the reaction-plane is identical within
uncertainties and also identical to the correlation observed in proton
+ proton collisions (solid lines). 
The away-side correlation is for in and out
of the reaction-plane suppressed compared to proton + proton
collisions, as expected in the case of jet quenching. 
It is also suggestive that the correlation out of the reaction plane
is suppressed more, which in case of jet quenching is expected due
to the different size of the system versus azimuth.

\begin{figure}[t]
  \begin{center}
    \includegraphics[width=.85\textwidth]{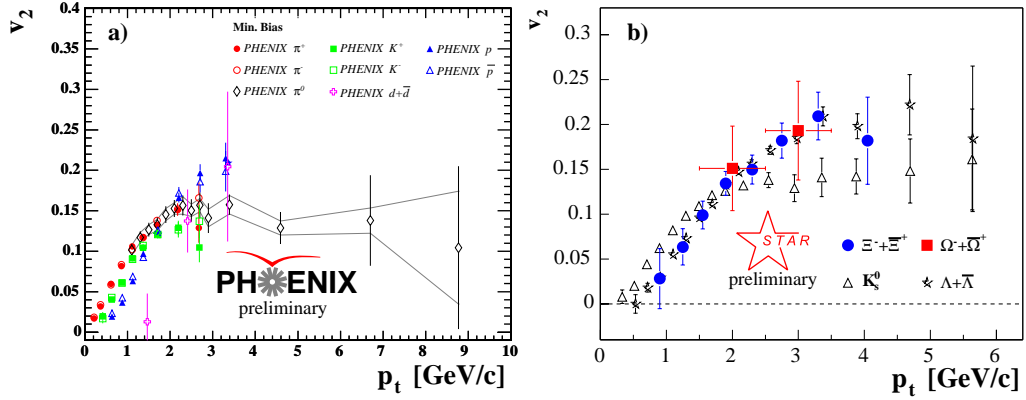}
  \end{center}
  \caption{
    a) Identified particle $v_2$ from PHENIX~\protect\cite{Kaneta:2004sb} 
    b) Identified particle $v_2$ from STAR~\protect\cite{Castillo:2004jy}
    \label{fig:identifiedflow}}
\end{figure}
Figure~\ref{fig:chargedflow}c shows the magnitude of the elliptic
flow, between 4 and 7 GeV/$c$, versus the collision centrality. 
The curves~\cite{Drees:2003zh} show various attempts, using different
energy loss scenarios, to describe 
the centrality dependence and the magnitude of $v_2$. 
The large magnitude of the
elliptic flow at high-$p_t$ is however difficult to interpret 
using the energy loss mechanism alone, as illustrated here by the
failure of the curves to describe the measurements. 
The measured $v_2$ is even
larger than expected in case of complete quenching~\cite{Shuryak:2001me}. 

A possible interpretation for the large magnitude of $v_2$ at
intermediate $p_t$ comes from the observed mass dependence.
Figures~\ref{fig:identifiedflow}a~\cite{Kaneta:2004sb}
and~\ref{fig:identifiedflow}b~\cite{Castillo:2004jy} show the $v_2$
for various particles as a function of $p_t$. At higher transverse
momenta the $v_2$ shows a particle dependence, which within
uncertainties, can be divided in two groups; baryons and mesons.
The proposed mechanism which naturally leads to a baryon/meson
scaling is constituent quark coalescence~\cite{Molnar:2003ff}. In this
scenario, at intermediate $p_t$, the mesons carry twice the
constituent quark $v_2$ (the baryons three times) which implies a
constituent quark $v_2$ of about 0.065~\cite{Sorensen:2004ij}. In the
suppression of particle production a baryon/meson scaling is also
observed. 
In the coalescence/recombination interpretation this follows
consistently~\cite{Fries:2004ej}.

\subsection{Particles with low transverse momentum}

The observations at intermediate and high-$p_t$ at RHIC clearly show
evidence of strong final state interactions.
One of the central questions which remains is if the
created system thermalizes, and if this happens early (during the
assumed partonic phase) in the collision. 
\begin{figure}[t]
  \begin{center}
    \includegraphics[width=.83\textwidth]{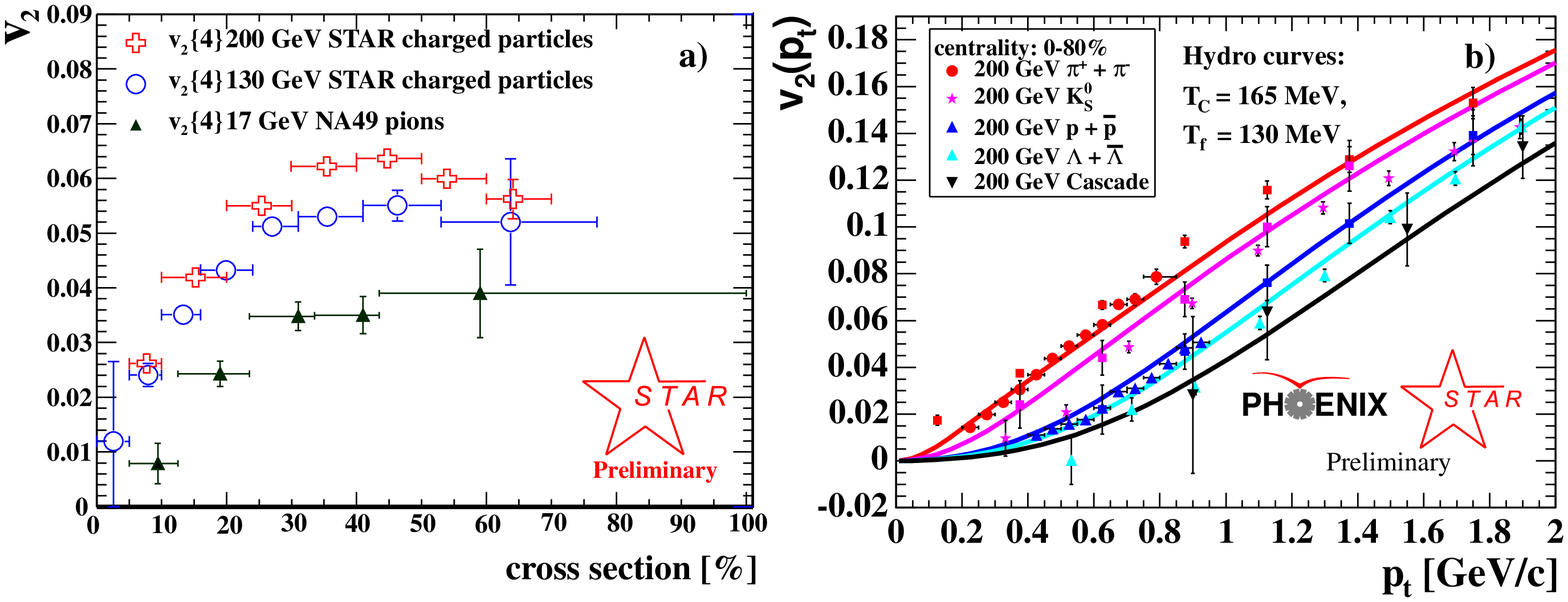}
  \end{center}
  \caption{
    a) Integrated elliptic flow values for the highest SPS energy,
    NA49~\protect\cite{Alt:2003ab}, and the
    two highest RHIC energies.
    b) Elliptic flow as a function of transverse momentum for various
    particles at low-$p_t$.
    \label{fig:identifiedflowlowpt}}
\end{figure}
One of the first observations at RHIC,
Fig.~\ref{fig:identifiedflowlowpt}a, showed a large increase in the
integrated elliptic flow from the highest SPS to the highest RHIC
energies, approaching the values predicted by ideal hydrodynamical
calculations which assumes local thermal equilibrium already after 0.2
fm/$c$. The collective behavior, one of the key features of a
hydrodynamical description, is particularly clear in the mass
dependence of the elliptic flow as shown in
Fig.~\ref{fig:identifiedflowlowpt}b. The hydrodynamical model
calculations~\cite{PasiPriv} give a good description of the elliptic
flow from the light pions to heavier particles like the $\Xi$.   

\section{Conclusions}

The observed strong collective flow and the strong suppression of 
high-$p_t$ hadrons indicates that a dense strongly interacting system
is created. The
system appears to a good approximation in local thermal equilibrium. 
The bulk of the system responds as a near-ideal, strongly coupled
fluid with an equation of state consistent with lattice QCD
calculations. There remain however important fundamental open
questions. Perhaps the most important one is the microscopic mechanism
underlying the apparent rapid thermalization. High precision
measurements of low-$p_t$ multi-strange and charmed hadron spectra and
correlations at RHIC and first results from the heavy-ion program at
the LHC will provide an important confirmation or perhaps new insights
to our current understanding.

\end{document}